\documentclass[aps,prd,floatfix,showpacs,twocolumn,superscriptaddress,nofootinbib]{revtex4-1}
\usepackage{dsfont}
\usepackage{amsmath,pslatex}
\usepackage{amsfonts}
\usepackage{amssymb}
\usepackage{graphicx}%
\usepackage{subfig}
\usepackage{braket}
\usepackage{float}
\usepackage{slashed}
\usepackage{hyperref}
\usepackage{color}
\usepackage{tensor}
\definecolor{darkgreen}{rgb}{0,0.35,0}

\newcommand{\p}{\partial}

\newcommand{\dd}{\ensuremath{\mathrm{d}}}

\newcommand{\be}{\begin{equation}}
\newcommand{\ee}{\end{equation}}
\newcommand{\bea}{\begin{eqnarray}}
\newcommand{\eea}{\end{eqnarray}}

\newcommand{\kulak}{KU Leuven Campus Kortrijk -- Kulak, Department of Physics, Etienne Sabbelaan 53 bus 7657, 8500 Kortrijk, Belgium}
\newcommand{\ughent}{Ghent University, Department of Physics and Astronomy, Krijgslaan 281-S9, 9000 Gent, Belgium}
\newcommand{\unesp}{Instituto de F\'{i}sica Te\'{o}rica, Universidade Estadual Paulista, Rua Dr.~Bento Teobaldo Ferraz, 271 - Bloco II, 01140-070 S\~{a}o Paulo, SP, Brazil}
\newcommand{\ucharles}{Faculty of Mathematics and Physics, Charles University, V Hole\v{s}ovi\v{c}k\'ach 2, 18000 Prague 8, Czech Republic}

\begin{document}

\title{On the exact quantum scale invariance of three-dimensional reduced QED theories}

\author{David Dudal}\email{david.dudal@kuleuven.be}\affiliation{\kulak}\affiliation{\ughent}
\author{Ana J\'{u}lia Mizher}\email{ana.mizher@kuleuven.be}\affiliation{\kulak}\affiliation{\unesp}
\author{Pablo Pais}\email{pablo.pais@kuleuven.be}\affiliation{\kulak}\affiliation{\ucharles}

\begin{abstract}
\noindent An effective quantum field theory description of graphene in the ultra-relativistic regime is given by reduced QED aka.~pseudo QED aka.~mixed-dimensional QED. It has been speculated in the literature that reduced QED constitutes an example of a specific class of hard-to-find theories: an interacting CFT in more than two dimensions. This speculation was based on two-loop perturbation theory. Here, we give a proof of this feature, namely the exact vanishing of the $\beta$-function, thereby showing that reduced QED can effectively be considered as an interacting (boundary) CFT, underpinning recent work in this area. The argument, valid for both two-~and four-component spinors, also naturally extends to an exactly marginal deformation of reduced QED, thence resulting in a non-supersymmetric conformal manifold. The latter corresponds to  boundary layer fermions between two different dielectric half-spaces.

\end{abstract}



\maketitle

Conformal invariance has played an important role in condensed matter physics and also high energy physics since the 1980's, in particular after the ground breaking work in $d=2$ dimensions of \cite{Belavin:1984vu} and its paramount relevance for string theory (world sheet dynamics). Establishing conformal invariance in $d>2$ turns out to be a much harder job, in the sense that not many examples are known of interacting (non-supersymmetric) conformal field theories (CFT) in $d>2$, see \cite{Nakayama:2013is,Baggio:2017mas} or \cite{Bashmakov:2017rko} for a few known examples and discussion.

In a recent work, it was investigated and proposed that mixed-dimensional Quantum Electrodynamics (QED) is another interacting (boundary) CFT \cite{Herzog:2017xha}, see also \cite{Karch:2018uft}. It arose in the context of new physics related to introducing a boundary into a CFT, in particular the appearance of extra boundary-related anomalous terms in the energy-momentum trace/correlation functions, and the latter connection with the standard anomaly contributions. One considers a four-dimensional bulk Abelian gauge field with action
\begin{eqnarray}\label{qed}
  \mathcal{S}_{\text{QED}_4} &=& \int \dd^4x \left[-\frac{1}{4}F_{\mu\nu}F^{\mu\nu}+ e j_\mu A^\mu\right]+S_{gf}\,,
\end{eqnarray}
coupled to three-dimensional (massless four-component) Dirac fermion matter via the conserved currents
\begin{eqnarray}\label{current}
  j^{\mu} &=& \left\{\begin{array}{cc}
              i\bar \psi \gamma^\mu \psi \delta(x_3) &  ~\text{for}~\mu=0,1,2\,, \\
              0& ~\text{for}~\mu=3\,,
            \end{array}\right.
\end{eqnarray}
with the fermion fields living on the boundary sheet $x_3=0$. Fermion dynamics can be included by adding the kinetic contribution $\int \dd^3x \bar\psi i \slashed{\partial}\psi$ to the system. As originally discussed \cite{Marino:1992xi,Gorbar:2001qt}, upon integrating out the four-dimensional bulk gauge field, followed by an integration over the third spatial direction orthogonal to the boundary plane, one ends with a non-local but fully three-dimensional gauge theory, which reads\footnote{From here one, we switched to Euclidean conventions.}
\begin{eqnarray}\label{rqed}
\mathcal{S}_{\text{RQED}_3}=\int \dd^3 x \left[ \frac{1}{2}   F_{\mu\nu} \frac{1}{\sqrt{-\partial^2}} F_{\mu\nu}+\bar{\psi} i\slashed{D}\psi  \right]+S_{gf}
\end{eqnarray}
after the introduction of a novel, but now three-dimensional, gauge field, that with a slight abuse of notion we denoted by $A_\mu$ again. As noted in \cite{Dudal:2018mms}, the gauge fixings in \eqref{qed}-\eqref{rqed} can be chosen independently, this is obviously due to the gauge invariant nature of the whole setup. The precise nature of the gauge fixing choice will be of little concern in the current note.

This version of mixed-dimensional QED, also known as Reduced QED (RQED$_3$) or Pseudo QED \cite{Nascimento:2015ola,Marino:2015uda}, already made its appearance in the literature before, as its physical relevance is motivated from condensed matter. Indeed, an effective quantum field theory description of the $\pi$ electrons in graphene, a two-dimensional plane of honeycomb ordered carbon atoms, is exactly provided by massless four-component Dirac spinors restricted to a plane, which evidently still interact through virtual photons than can propagate in the four-dimensional surrounding bulk \cite{CastroNeto:2009zz,Gusynin:2007ix,Vozmediano:2010zz,Gonzalez:1993uz}. The unitarity of the unusual looking theory \eqref{rqed} was established recently in \cite{Marino:2015uda}. Strictly speaking, for graphene, the 2D $\vec\nabla$-operator inside the $\slashed{D}$ is to be replaced by $\frac{v_F}{c} \vec\nabla\approx \frac{\vec\nabla}{300}$ with $v_F$ the Fermi velocity, but here we will consider the Lorentz invariant version, that is with $v_F=c$, the speed of light in vacuum.

Although RQED$_3$ as described by the action \eqref{rqed} looks very similar to QED$_3$, there is one crucial difference. The electromagnetic coupling $e^2$ is still dimensionless now, since it originates from the four-dimensional standard gauge interaction\footnote{This can also be easily confirmed from the action \eqref{rqed} by classical power counting of dimensions.}, while in the three-dimensional case the coupling carries an intrinsic dimension. The theory, for massless Dirac fermions, is thus classically scale invariant. Two-loop computations, \cite{Teber:2012de,Teber:2018goo} revealed that the coupling $e^2$ does not run, i.e.~it does not get renormalized in a massless renormalization scheme like $\overline{\mbox{MS}}$. A similar one-loop observation in the context of graphene was made in \cite{Vozmediano:2010fz,Gonzalez:1993uz} and up to second order in \cite{Hands:1994kb} for what concerns the Thirring model in a large $N_f$ expansion.

A non-relativistic version, for $N$ species of two-component spinors, of the model \eqref{rqed} was introduced and analyzed in \cite{Son:2007ja}, also leading to the question whether the theory is scale invariant (conformal invariant\footnote{We must note here that, from a strictly mathematical point of view, scale invariance is a weaker condition than conformal invariance \cite{DiFrancesco_book}. In $d=2$ dimensions it was proven that scale invariance implies conformal invariance \cite{Polyakov:1970xd}. However, once scale invariance is determined, a sufficient condition for conformal invariance is attainable in $d>2$ dimensions, providing the non-existence of an integrated operator transforming as a vector under rotations with scale dimension $-1$ \cite{Polchinski1988}. A similar condition was proposed for the three-dimensional Ising model, see \cite{Delamotte:2015aaa}, and more recently \cite{DePolsi2018}.}) or not at finite $N$, in relation to the phase structure: can a gap open or not? Even for genuine QED$_3$ this question is still under debate, \cite{Appelquist:1986fd} reported a dynamical gap for sufficiently small $N$ while recent lattice studies \cite{Karthik:2016ppr} found no evidence of such for $N=2$.

Returning to RQED$_3$, the authors of \cite{Herzog:2017xha} motivated for the coupling $e^2$ to be an all orders fixed point of the renormalization group equation, i.e.~RQED$_3$ would be an example of an interacting non-supersymmetric CFT, defining an at least perturbatively stable conformal manifold as designated in \cite{Bashmakov:2017rko} upon inclusion of an electromagnetic interaction that ``jumps'' across the boundary $x_3=0$, as considered in \cite{Karch:2018uft}.  We will come back to this latter model later on. CFT aspects of RQED were also highlighted in \cite{Menezes:2016euw}.

The goal of the current paper is to give an affirmative answer to the above. To be more precise, we will show that the $\beta$-function for the RQED$_3$ coupling $e^2$ is exactly vanishing in massless renormalization schemes, including the case with the above deformation. Let us mention that for standard QED$_3$, with its massive coupling $e^2$, the complete IR and UV finiteness was proven in \cite{DelCima:2013gpa} using the BPHZL framework. Notice that in \cite{Hands:1994kb}, a similar line of reasoning was employed to motivate the renormalizability (not finiteness!) of the Thirring model at large $N_f$.

Let us depart from the would-be bare action in $d=3-\epsilon$ dimensions,
\begin{eqnarray}\label{rqedb}
\mathcal{S}_{\text{RQED}_3}&=&\int \dd^{3-\epsilon} x \left[ \frac{1}{2}Z_A^2   F_{\mu\nu} \frac{1}{\sqrt{-\partial^2}} F_{\mu\nu}+Z_\psi\bar{\psi} i\slashed{\p} \psi\right.\nonumber\\&& \left.+ Z_\Gamma \bar\psi i\slashed{A}\psi   \right]+S_{gf}\,,
\end{eqnarray}
that is, including all renormalization $Z$-factors for the photon field $A$, the fermion fields $(\psi,\bar\psi)$ and the fermion-photon vertex. Just as for normal QED$_4$, current conservation translates into a Ward identity \cite{Ryder:1985wq}, linking the 1$PI$ fermion-photon vertex $\Gamma_\mu^{(3)}$ to the inverse (1$PI$) fermion propagator $\Gamma_\mu^{(2)}$,
\begin{eqnarray}\label{Z1}
  q_\mu \Gamma_\mu^{(3)}(p,q,p+q)=\Gamma^{(2)}(p+q)-\Gamma^{(2)}(p)\,,
\end{eqnarray}
or, taking $q_\mu\to0$,
\begin{eqnarray}\label{Z1bis}
  \Gamma_\mu^{(3)}(p,0,p)=\frac{\p\Gamma^{(2)}(p)}{\p p_\mu}\,.
\end{eqnarray}
At the level of the earlier $Z$-factors, this translates into $Z_\Gamma=Z_\psi$, from which it then follows that
\begin{equation}\label{Z2}
  e^2= \mu^{-\epsilon} Z_A e_0^2
\end{equation}
with $e_0$ the bare charge. So in principle it is sufficient to prove the finiteness of the photon renormalization factor $Z_A$ to have $\beta_{e^2}=0$ for $\epsilon\to0$. Considering the 1$PI$ photon propagator (self-energy) $\Pi_{\mu\nu}(p^2)$, power counting leads to superficial degree of divergence $\nu$ at $n$-loops \cite{Teber:2012de,Herzog:2017xha}, namely $\nu = 1$. As in general, gauge (or better said BRST) invariance imposes the photon self-energy to be transverse, one can factor out a $\delta_{\mu\nu} p^2-p_\mu p_\nu$ from $\Pi_{\mu\nu}(p^2)$, leading to a superficially convergent diagram.  Unfortunately, this argument, used at one-loop in \cite{Herzog:2017xha}, does not help at a generic order, since (i) there will be a sum of diagrams contributing to $\Pi_{\mu\nu}(p^2)$ with only the sum transverse and (ii), any higher order diagram is superficially convergent if and only all of its subdiagrams are \cite{Weinberg:1959nj,Zimmermann:1968mu}, and the latter subdiagrams also do not need to be transverse by themselves.

In \cite{Teber:2018goo}, it was pointed out that $Z_A=1$ as it concerns the renormalization of a non-local term in the free (quadratic) part of action, incompatible with the observation that counterterms must be local polynomials in the fields and derivatives thereof. This rationale was based on \cite{Vasilev:2004yr}. However, the argument of \cite{Vasilev:2004yr} is based on adding on top of a renormalizable theory a non-local term. For example, consider
\begin{equation}\label{phi4a}
  S=\int \dd^4x\left[-\frac{1}{2} \phi \left(\p^2+\frac{m^4}{\p^2}\right)\phi+\frac{\lambda}{4!}\phi^4\right]\,,
\end{equation}
then this theory is a standard local renormalizable quantum field theory for $m^4=0$, and it remains to be so when the dipole term $\propto m^4$ is switched on; indeed the only change is the propagator, now given by\footnote{A similar partial fraction trick was used in \cite{Capri:2015mna} in a different context.} $\frac{p^2}{p^4+m^4} = \frac{1}{p^2}-\frac{m^4}{p^2(p^4+m^4)}$, and the second ultraviolet suppressed term will not generate new infinities compared to the first original piece of the propagator. As such, no counterterm for the dipole piece of the action is necessary. The crux of the matter here is that the underlying (local) quantum field theory is already properly renormalized.  The situation however changes drastically if there is no such underlying renormalizable theory. Consider for example
\begin{equation}\label{phi4b}
  S=\int \dd^4x\left[-\frac{1}{2} \phi \left(\frac{\p^2}{\sqrt{-\p^2}}\right)\phi+\frac{\lambda}{4!}\phi^4\right]\,.
\end{equation}
Dimensional counting learns that $\lambda$ has negative mass dimension.  As such, we do not expect this model to be renormalizable to all orders.  Apart from that, the  ``setting sun'' self energy diagram will anyhow require wave function renormalization, visible per power counting. The problem of course is that the free $\phi$-propagator now only falls off like $\frac 1 p$ in the ultraviolet. Moreover, the fact that counterterms are polynomials in the momentum has strictly speaking only be proven when using free propagators of the standard type, see \cite{Zimmermann:1969jj,Lowenstein:1975ps,Breitenlohner:1975hg,Breitenlohner:1976te}.

Therefore, another technology is needed to prove that $Z_A=1$. Let us start with the action \eqref{rqedb} and integrate out the fermions \`{a} la \cite{Barci:1998zd}, to get an effective theory for photons only, from which we can also read off the $Z_A$.  Integrating out the fermions leads to
\begin{eqnarray}\label{rqedc}
\tilde\Gamma[A]&=&\int \dd^{3-\epsilon} x \left[ \frac{1}{2}Z_A^2   F_{\mu\nu} \frac{1}{\sqrt{-\partial^2}} F_{\mu\nu}\right]\nonumber\\&&+\ln \det (i\slashed{D})  +S_{gf}\,,
\end{eqnarray}
where $A$ is here considered to be still external\footnote{This determinant and the emergent Chern--Simons term plays an important r\^{o}le in 3D bosonization and dualities, see \cite{Banerjee:1995ry,Schaposnik:1995np,Barci:1995iy,Barci:1998zd,Deser:1981wh,Deser:1982vy,Redlich:1983dv}. Recently there has been an revived  activity in such dualities, in particular in relation to {\sf T}-invariance and two-component spinor theories, an interest sparked by papers like \cite{Son:2015xqa,Karch:2016sxi,Seiberg:2016gmd}. To avoid confusion, although we relied on tools known in the bosonization community, we do \emph{not} derive a dual version of the four-component spinor theory RQED$_3$. The four-component nature of our spinors makes that the theory \eqref{rqed} is not prone to a $\sf T$-parity anomaly. Moreover, thinking in terms of graphene, the four-component language automatically emerges. Indeed, the honeycomb lattice structure of graphene actually consists out of two periodic sublattices as which creation/annihilation operators can be inserted, leading to two Dirac points in the momentum space, and the expansion around these points  can be managed to form a four-spinor structure in the continuum limit \cite{Gusynin:2007ix,Jackiw:2007rr}.
}. Gauge symmetry translates now into
\begin{eqnarray}\label{rqedc2}
\p_{\mu_1}\frac{\delta\tilde\Gamma[A]}{\delta A_{\mu_1}}&=&0\,.
\end{eqnarray}
Taking further functional derivatives w.r.t.~$A_i\equiv A_{\mu_i}(x_i)$ and setting external fields to zero at the end, we get
\begin{eqnarray}\label{rqedc2}
\left.\p_{\mu_1}^{x_1}\frac{\delta^{(n)}}{\delta A_1\ldots\delta A_n}\tilde\Gamma[A]\right|_{A=0}&=&\p_{\mu_1}^{x_1}\braket{ j_{\mu_1}^{x_1}\ldots j_{\mu_n}^{x_n}}=0\,,
\end{eqnarray}
expressing that $\tilde\Gamma(A)$ is actually solely depending on the transverse projection of $A$, viz.~$\tilde\Gamma(A)=\tilde\Gamma(A^T)$
where
\begin{eqnarray}\label{rqedc4}
A_\mu^T=\left(\delta_{\mu\nu}-\frac{\p_\mu \p_\nu}{\p^2}\right)A_\nu\,.
\end{eqnarray}
This non-local variable $A_\mu^T$ is gauge invariant, so unsurprisingly, we can rewrite it in terms of $F_{\mu\nu}$ via ($d=3$)
\begin{eqnarray}\label{rqedc5}
A_\nu^T= \frac{\p_\mu}{\p^2}F_{\mu\nu}=\int \frac{\dd^3r}{4\pi}\frac{(x-r)_\mu}{|x-r|^3}F_{\mu\nu}^{r}\,.
\end{eqnarray}
Next, we consider the all-order expansion of $\tilde \Gamma(A)$, being
\begin{eqnarray}\label{rqedc6}
\tilde\Gamma&=&\sum_{n\geq1} \int \dd^3x_1\ldots \dd^3 x_n A_1^T\ldots A_n^T\braket{j_{\mu_1}^{x_1}\ldots j_{\mu_n}^{x_n}}\\
&=&\sum_{n\geq1} \int \dd^3r_1 \ldots \dd^3r_n F_{\mu_1 \nu_1}^{r_1}\ldots F_{\mu_n \nu_n}^{r_n} \gamma_{\mu_1\nu_1,\ldots, \mu_n\nu_n}^{r_1,\ldots, r_n}\nonumber
\end{eqnarray}
with
\begin{eqnarray}\label{rqedc7}
 \gamma_{\mu_1\nu_1,\ldots, \mu_n\nu_n}^{r_1,\ldots, r_n}&=& \int \frac{\dd^3x_1}{4\pi} \ldots \frac{\dd^3x_n}{4\pi}\frac{(x_1-r_1)_{\mu_1}}{|x_1-r_1|^3}\ldots\nonumber\\&&\times\frac{(x_n-r_n)_{\mu_n}}{|x_n-r_n|^3}\braket{ j_{\nu_1}^{x_1}\ldots j_{\nu_n}^{x_n}}\,.
\end{eqnarray}
As charge conjugation invariance is also valid in three dimensions and its operation switches the sign of the current, Furry's theorem still holds and we will just encounter the even terms in the expansion \eqref{rqedc6}. It is easy to see that a diagram with $n$ external photon legs will behave in the ultraviolet as $\sim \int\dd^3q\frac{1}{q^{n}}$, so we need to only consider the $n=2$ case for possible divergences, the other diagrams are power-counting finite in $d=3$, as $n\geq 4$. The two-current expectation value is nothing else than the transverse photon self-energy for which a standard computation for a single four-component spinor, see also \cite{Gorbar:2001qt,Barci:1995iy}, leads to a \emph{finite} correction in $d=3-\epsilon$ dimensions
\begin{eqnarray}\label{rqedc8}
 \Pi_{\mu\nu}(p)=\frac{e^2}{8 p}\left(\delta_{\mu\nu}-\frac{p_\mu p_\nu}{p^2}\right)\,.
\end{eqnarray}
Putting everything back together, we will get as effective action for the photon in RQED$_3$
\begin{eqnarray}\label{rqedd}
\tilde\Gamma[A]&=&\int \dd^{3-\epsilon} x \left[ \frac{1}{2}Z_A^2   F_{\mu\nu} \frac{1}{\sqrt{-\partial^2}} F_{\mu\nu}  \right.\\\nonumber&&+\left.\frac{e^2}{8}F_{\mu\nu} \frac{1}{\sqrt{-\partial^2}} F_{\mu\nu}+\mathcal{O}\left(\frac{e^4F^4}{\sqrt{-\p^2}^5}\right)  \right]+S_{gf},
\end{eqnarray}
From this expression, it is clear that the effective interactions in the higher powers of the field strength $F$ are sufficiently  ultraviolet-suppressed to only give power counting finite corrections, as such it is evident that we can actually set $Z_A\equiv1$, what we wanted to prove. To make this explicit, consider e.g.~the vertex $\sim \frac{F^4}{p^5}$ and consider a diagram with $N\geq 2$ external legs\footnote{Vacuum diagrams in massless theories or one-point propagators are vanishing anyhow.} and $V\geq1$ vertices. For a number of $L$ loops we have $L=P-V+1$, with $P$ the number of propagators. Each vertex counts 4 photon lines, hence $4V = N+2P$. Keeping in mind that the propagator falls off as $\frac{1}{p}$, the considered diagram will thus have a superficial degree of divergence given by $\nu=3L-P-3V=-2V+3-N<0$, i.e.~it will be convergent. A similar argument will apply if further UV suppressed vertices are included.

Having established the proof for the four-component case, it is in fact immediately realized that the same line of reasoning can be followed in case the fermion is two-component. Indeed, the only change, up to the replacement  $\frac{e^2}{8}\to\frac{e^2}{16}$, in \eqref{rqedd} will be the additional generation of a (finite)  ${\sf T}$-odd Chern--Simons term $\propto \int \dd^3x\left(e^2\epsilon_{\mu\nu\rho}A_\mu \p_\nu A_\rho\right)$ which also respects gauge invariance \cite{Deser:1981wh,Deser:1982vy,Redlich:1983dv,Barci:1995iy,Schaposnik:1995np}. Said otherwise, one still finds that $Z_A=1$.

Notice that, silently, we assumed during the above line of reasoning that the fermions have a Fermi velocity $v_F=c$ with $c$ the speed of light in the layer, i.e.~to have full 3D Lorentz (Euclidean) invariance. Though, in a realistic condensed matter system, we should take into account the fermions having a Fermi velocity $v_F<c$. This is a highly non-trivial addition to the setup, since $v_F$ generically renormalizes (see e.g.~\cite{Vozmediano:2010fz,Gonzalez:1993uz,Stauber:2017fuj} for theoretical considerations or \cite{diraccones2,diraccones} for experimental evidence), which indirectly also causes the interaction to run since the effective ``fine structure constant'' is given by (restoring all units) $\frac{e^2}{4\pi \hbar v_F}$ \cite{Gonzalez:1993uz}. Though, the Lorentz invariant CFT description should be effectively realized in the low energy limit, where $v_F$ runs to the infrared fixed point $v_F=c$, viz.~the Lorentz invariant case \cite{Vozmediano:2010fz,Gonzalez:1993uz,Stauber:2017fuj}.

We can now move to a further generalization of our setup by looking at the theoretical model of \cite{Karch:2018uft}, which we generalize further by considering
\begin{eqnarray}\label{karch1}
  S_{ins} &=& \int  \dd t\dd^3x \left[\frac{\theta(x_3)}{4e_+^2}F_{\mu\nu,+}^2+   \frac{\theta(-x_3)}{4e_-^2}F_{\mu\nu,-}^2\right]+S_{gf} \nonumber\\&&  +\int \dd t\int\dd^2x \left[\bar{\psi} i\slashed{D}\psi\right]
\end{eqnarray}
where $\theta(x)$ is the Heaviside step function. We introduced $A_{\mu,\pm}$ for the gauge field values above/below the $x_3=0$ boundary plane, with $\left[A_\mu\right]_{x_3=0}=\frac{1}{2}\left[a_{\mu,+}A_{\mu,+} + a_{\mu,-}A_{\mu,-}\right]_{x_3=0}$ where $a_{0,\pm} = \frac{c_\pm}{v_F}\,,~a_{1,\pm}=a_{2,\pm}=1$. It is understood that $\slashed{\p}=\gamma_0 \frac{1}{v_F}\frac{\p}{\p t}+ \vec{\gamma}\cdot\vec{\nabla}$ while current conservation is expressed via $\frac{1}{v_F}\partial_t j_0+ \vec{\nabla}\cdot \vec{j}=0$. As before, $j_\mu\equiv i\bar\psi \gamma_\mu \psi$ for $\mu=0,1,2$ and $j_3=0$, with $j_\mu$ not depending on $x_3$. The model is gauge invariant, in particular due to how $\left[A_\mu\right]_{x_3=0}$ is introduced. The setup corresponds to a surface layer of massless fermions between two different dielectric media (insulators). We allowed for a different speed of light in the two surrounding media ($c_+$ and $c_-$), so that $F_{\mu\nu,\pm}^2= E^2/c_\pm^2+B^2$, next to a different interaction strength, incorporated in the $e_+^2$ and $e_-^2$. The description \eqref{karch1} corresponds to a realistic model for an isotropic insulator \cite[Sect.~16.10]{Fradkin:1991nr}. The special case $c_+=c_- =v_F(=1)$ matches to the example given in \cite{Karch:2018uft} and this is the one we will be discussing here. The presented methodology can be adapted to the general case, although matters get considerably more tedious. In any case, as before we only expect the model to be scale anomaly free for $c_+=c_-=v_F$.

To prove that \eqref{karch1} enjoys an exact quantum scale invariance for $c_+=c_-=v_F=1$, we will first reduce it to a $3D$ model describing the interaction between the planar fermions. As the gauge field appears at most quadratically, we can integrate it out exactly, equivalent to working with the on-shell action. The classical field equations read $\p^2 A_{\mu,\pm}=0$ where we assumed Landau gauge $\p_\mu A_{\mu,\pm}=0$. There is an extra set of constraints as we must require the boundary variation to vanish as well. With $n_\mu=(0,0,0,1)$, this leads to
$\left[\frac{1}{e_+^2}n^{\mu} F_{\mu\nu,+}-\frac{1}{e_-^2}n^{\mu} F_{\mu\nu,-}\right]_{x_3=0}=j_{\nu}$. Moreover, requiring continuity of the Bianchi identity leads to $\left[n_\mu \epsilon^{\mu\nu\alpha\beta}(F_{\alpha\beta,+}-F_{\alpha \beta,-})\right]_{x_3=0}=0$, the homogenous boundary conditions.  Using a similar approach as in \cite{Blommaert:2018oue}, we can construct an explicit solution in terms of the Fourier-transformed current $\hat{j}_{\mu}$,
\begin{eqnarray}
  A_{\mu,\pm} &=& -\frac{e_+^2 e_-^2}{e_+^2 + e_-^2}\int \frac{\dd^3k}{(2\pi)^3}\frac{e^{i(k_0x_0+k_1x_1+k_2x_2)\mp k_{3}x_3}}{k_{3}}\hat j_\mu\quad \nonumber\\&\text{for}&~\mu=0,1,2~\text{and with} ~k_{3}= \sqrt{k_0^2+k_1^2+k_2^2}\,,\nonumber\\
  A_{3,+}&=&A_{3,-}\equiv0\,,
\end{eqnarray}
which is easily checked to fulfill the gauge condition, the field equations and the boundary conditions. The on-shell action becomes pure boundary, yielding
\begin{eqnarray}
S_{ins}&=& \int \dd^3 x  \frac{1}{2}\left[A_{\mu,+}+ A_{\mu,-}\right]_{x_3=0}j_\mu+\int \dd^3x \bar\psi i\slashed{\p}\psi+\\&&\hspace{-1cm}\left[-\frac{1}{2 e_+^2}\int \dd^3x A_{\mu,+} \p_3 A_{\mu,+}+ \frac{1}{2 e_-^2}\int \dd^3x A_{\mu,-} \p_3 A_{\mu,-}\right]_{x_3=0}\nonumber\\
&=&  -\frac{e_+^2 e_-^2}{e_+^2+e_-^2}\int \frac{\dd^3 k}{(2\pi)^3} \hat j_\mu \frac{1}{2k_3} \hat j_\mu+\int \dd^3x \bar\psi i\slashed{\p}\psi\nonumber\\
&=&  -\frac{e_+^2 e_-^2}{e_+^2+e_-^2}\int \frac{\dd^3 k}{(2\pi)^3} \hat j_\mu \frac{1}{2k_3}P_{\mu\nu} \hat j_\nu+\int \dd^3x \bar\psi i\slashed{\p}\psi\,,\nonumber
\end{eqnarray}
with $P_{\mu\nu}$ the $3D$ transverse projector. Returning to configuration space, we can reformulate the mixed-dimensional model \eqref{karch1} in terms of a new $3D$ gauge field via
\begin{eqnarray}\label{rqedtris}
S_{ins}&=&\int \dd^3 x   \left[\frac{1}{2 \tilde e^2}F_{\mu\nu}\frac{1}{\sqrt{-\p^2}}  F_{\mu\nu}+\bar{\psi} i\slashed{D}\psi\right] +S_{gf}
\end{eqnarray}
 with $\tilde e^2=\frac{2(e_+^2 e_-^2)}{e_+^2+e_-^2}$ the new effective $3D$ electromagnetic coupling. This means that the two coupling constants $e_\pm^2$ will never enter separately, but always in the combination $\tilde e^2$. As we recover RQED$_{3}$, see equation \eqref{rqed}, with appropriate coupling, we can still conclude that the $\beta$-function of $\tilde e^2$ is trivial, whatever the values of $e_\pm^2$. This proves the point made in \cite{Karch:2018uft}. As a check, in the case of two identical dielectrics, we recover the effective graphene model discussed earlier and derived in a different manner in e.g.~\cite{Marino:1992xi,Gorbar:2001qt}.

In separate work, we plan to come back to the original model with $c_\pm$ and $v_F$ present. A particular interesting question is whether by appropriate choices of $e^2_\pm$, $c_\pm$ and $v_F$, \mbox{(non-)}trivial fixed points can be found, and if so, to what extent these can be realized in Nature? We conclude by discussing in short possible experimental realizations of the above theoretical model. A first possibility is to consider a sheet of graphene between two different dielectrics \cite{dassarma}. Another interesting setup is to make use of the massless (chiral) fermions living on the three-dimensional edge between the insulating vacuum and a (3+1)-dimensional topological insulator, \cite{Qi:2008ew,Fradkin:1991nr}. Interestingly, in the latter case the four-dimensional description of the $\mathbb{Z}_2$ topological insulator has a topological $\propto\theta \int \dd^4x F\tilde F\propto \theta \int\dd^4x \vec{E}\cdot\vec{B}$ term in the action with $\tilde F_{\mu\nu}=\frac{1}{2}\epsilon_{\mu\nu\alpha\beta}F_{\alpha\beta}$ the dual field strength tensor, with the angular variable $\theta=\pi$ to respect $\sf T$-invariance. For the vacuum, we have $\theta=0$. Upon integration, this jump in $\theta$ will exactly produce the 3D Chern--Simons term on the boundary for the 3D dimensionally-reduced photon, since $\int \dd^4x F\tilde F\propto\int \dd^4x \epsilon_{\mu\nu\alpha\beta} \p_\mu (A_\nu \p_\alpha A_\beta)=\int \dd^3x \epsilon_{\nu\alpha\beta} A_\nu \p_\alpha A_\beta $ assuming $x_\mu\equiv x_3=0$ is the boundary. As such, topological insulators offer the possibility to explicitly couple the Chern--Simons photon term also to reduced QED, as recently discussed in \cite{Dudal:2018mms}, see also \cite{Coleman:1985zi}. At least in the Lorentz invariant limiting case, this 3D model will also have no $\beta$-function for the electromagnetic coupling, following the analysis in our current note.

At last, having shown that in the ultrarelativistic limit description of graphene there is no space for coupling constant renormalization, this also means that a priori massless fermions will never be able to generate a dynamical mass given that there is no space for dimensional transmutation with a vanishing $\beta$-function. This can be circumvented by introducing external dimensionfull quantities (like background electromagnetic fields) or by taking into account that realistic graphene has a natural ultraviolet cut-off inversely proportial to the cell size. These and other issues deserve further attention in future research.

\section*{Acknowledgments}
We thank S.P.~Sorella for most useful discussions during the preparation of this work, as well as N.~Bobev and M.~N.~Chernodub for some helpful comments. A.~Mizher is a beneficiary of a postdoctoral grant of the Belgian Federal Science Policy (BELSPO) and receives partial support from FAPESP under fellowship number  2016/12705-7, while the work of P.~Pais is supported by a PDM grant of KU Leuven.

\bibliography{RQEDbeta_biblio}{}
\bibliographystyle{apsrev4-1}

\end{document}